\documentclass[aps,floatfix,showpacs,twocolumn]{revtex4}
\usepackage{graphicx}

\begin{document}

\title{All static spherically symmetric perfect fluid solutions of Einstein's Equations}
\author{Kayll Lake \cite{email}}
\affiliation{Department of Physics, Queen's University, Kingston,
Ontario, Canada, K7L 3N6 }
\date{\today}

\begin{abstract}
An algorithm based on the choice of a single monotone function
(subject to boundary conditions) is presented which generates all
regular static spherically symmetric perfect fluid solutions of
Einstein's equations. For physically relevant solutions the
generating functions must be restricted by non-trivial
integral-differential inequalities. Nonetheless, the algorithm is
demonstrated here by the construction of an infinite number of
previously unknown physically interesting exact solutions.
\end{abstract}


\maketitle

Exact solutions of Einstein's field equations provide a route to
the physical understanding (and discovery) of relativistic
phenomena, a convenient basis from which perturbation methods can
proceed and a check on numerical approximations. Here we look at
static spherically symmetric perfect fluid solutions.
Unfortunately, even for this simple type, very few solutions are
in fact known, and of these few pass even elementary tests of
physical relevance \cite{dellake}. In this paper, an algorithm
based on the choice of a single monotone function (subject to
boundary conditions) is presented which generates all regular
static spherically symmetric perfect fluid solutions of Einstein's
equations. We are interested only in physically relevant solutions
here and so the algorithm must be supplemented by physical
considerations \cite{physical}. These additional conditions limit
the generating functions allowed by way of non-trivial
integral-differential inequalities. The details of how to choose
physically relevant generating functions (beyond trial and error)
are, at present, not known. Nonetheless, the robustness of the
algorithm is demonstrated here by the construction of an infinite
number of previously unknown physically interesting exact
solutions.

\bigskip
To set the notation, consider a spherically symmetric spacetime
$\mathcal{M}$  \cite{metric}
\begin{equation}
ds^2_{\mathcal{M}}=ds^2_{\Sigma}+R^2d\Omega^2 \label{ss}
\end{equation}
where $d\Omega^2$ is the metric of a unit sphere
($d\theta^2+sin^2(\theta)d\phi^2$) and $R=R(x^1,x^2)$ where the
coordinates on the Lorentzian two space $\Sigma$ are labelled as
$x^1$ and $x^2$.  Consider a flow (a congruence of unit timelike
vectors $u^{\alpha}$) tangent to an open region of $\Sigma$ and
write $n^{\alpha}$ as the normal to $u^{\alpha}$ in the tangent
space of $\Sigma$. Both $u^{\alpha}$ and $n^{\alpha}$ are uniquely
determined. We suppose that (\ref{ss}) is generated by a fluid
subject to the condition $G_{\alpha}^{\beta}u^{\alpha}n_{\beta}=0$
where $G_{\alpha}^{\beta}$ is the Einstein tensor (see
\cite{lake}). Let $G \equiv  G_{\alpha}^{\alpha}, \; G1 \equiv
G_{\alpha}^{\beta}u^{\alpha}u_{\beta}$ and $G2 \equiv
G_{\alpha}^{\beta}n^{\alpha}n_{\beta}$.  In the static case it
follows that the flow is shear free and that
\begin{equation}
G+G1=3G2 \label{condition}
\end{equation}
is a necessary and sufficient condition for (\ref{ss}) to
represent a perfect fluid \cite{TOV}.

\bigskip
First consider $\mathcal{M}$ in ``curvature" coordinates,
\begin{equation}
ds^2_{\mathcal{M}}=\frac{dr^2}{1-\frac{2m(r)}{r}}+r^2d\Omega^2-e^{2\Phi(r)}dt^2.
\label{standardform}
\end{equation}
Writing out (\ref{condition}) \cite{condition} we obtain an
expression involving $\Phi(r)$ and $m(r)$ with derivatives to
order two in $\Phi(r)$ and to order one in $m(r)$. Viewing
(\ref{condition}) as a differential equation in $\Phi(r)$, given
$m(r)$, we obtain a Riccati equation in the first derivative of
$\Phi(r)$. However, viewing (\ref{condition}) as a differential
equation in $m(r)$, given $\Phi(r)$, we obtain a linear equation
of first order \cite{berger}. As a consequence, we have the
following algorithm for constructing all possible spherically
symmetric perfect fluid solutions of Einstein's equations:

\bigskip
Given $\Phi(r)$ (sufficiently smooth and subject to boundary
conditions explained below)
\begin{equation}
m(r) =\frac{\int \!b ( r) {e^{\int \!a ( r)
{dr}}}{dr}+\mathcal{C}}{{e^{\int \!a(r) {dr}}}} \label{mass}
\end{equation}
where
\begin{equation}
a(r) \equiv{\frac {2\,   {r}^{2}(\Phi^{''} ( r )+ \Phi ^{'}( r )
^{ 2})
 -3 r\,  \Phi^{'} ( r )
 \, -3}{r (  r \Phi^{'} ( r )
  +1 ) }}\label{a}
\end{equation}
and
\begin{equation}
b(r) \equiv{\frac {r ( r (\Phi^{''} ( r
 )   +  \Phi ^{'}( r )
  ^{2}) -\Phi^{'} ( r )  ) }{
 r \Phi^{'} ( r )   +1}}\label{b}
\end{equation}
with $^{'}\equiv\frac{d}{dr}$ and $\mathcal{C}$ a constant. The
generating function associated with any known solution is of
course immediately obvious following the algorithm.

\bigskip
Interior boundary conditions on $\Phi(r)$ are set by the
requirement that all invariants polynomial in the Riemann tensor
are finite at the origin. In this case there are but three
independent invariants \cite{invariants} and these are expressed
here in terms of the physical variables; the energy density
\begin{equation}
\rho=\frac{G1}{8 \pi}=\frac{m^{'}(r)}{4 \pi
r^2}\geq0\label{energy}
\end{equation}
and the isotropic pressure
\begin{equation}
 p=\frac{G2}{8 \pi}=\frac{r
\Phi^{'}(r)(r-2m(r))-m(r)}{4 \pi r^3}\geq0.\label{pressure}
\end{equation}
 Note that the inequalities in (\ref{energy}) and (\ref{pressure}) are to be viewed as
imposed restrictions on $\Phi(r)$. At the centre of symmetry
($r=0$) the regularity of the Ricci invariants requires that
$\rho(0)$ and $p(0)$ be finite. The regularity of the Weyl
invariant requires that  $m(r)$ is $C^3$ at $r=0$ with
$m(0)=m(0)^{'} = m(0)^{''} = 0$ and $m(0)^{'''}=8 \pi \rho(0)$
\cite{schw}. In summary, for a static spherically symmetric
perfect fluid, finite $\rho(0)$ and $p(0)$ guarantees the
regularity of all Riemann invariants at the centre of symmetry.
$\Phi(0)$ is a finite constant (set by the scale of $t$) and it
follows from (\ref{pressure}) that  $\Phi^{'}(0)=0$ and
$\Phi^{''}(0)=\frac{4 \pi}{3}(3p(0)+\rho(0))>0$. Since $\rho\geq0$
and continuous and since $p(0)>0$ and finite it follows from
(\ref{condition}) that $r>2m(r)$ \cite{br}. With $r>2m(r)$ for
$r>0$ it also follows from (\ref{pressure}) for $p(r)>0$ that
$\Phi^{'}(r)\neq 0$ for $r>0$. As a result, the source function
$\Phi(r)$ must be a monotone increasing function with a regular
minimum at $r=0$. Exterior boundary conditions on $\Phi(r)$ exist
only for isolated spheres and these conditions are set by junction
conditions \cite{junctions}. The necessary and sufficient
condition that ${\mathcal{M}}$ have a regular boundary surface
with a Schwarzschild vacuum exterior at $r=R>0$ is given by
$p(r=R)=0$. Setting $m(r=R) \equiv M$ it follows that
$\Phi^{'}(r=R)=\frac{M}{R(R-2M)}$.

\bigskip
Each source function $\Phi(r)$ which is a monotone increasing
function with a regular minima at $r=0$ necessarily gives, via
(\ref{mass}), a static spherically symmetric perfect fluid
solution of Einstein's equations which is regular at $r=0$. Exact
solutions, in the present context, can be viewed as those for
which (\ref{mass}) can be evaluated without recourse to numerical
methods. The number of source functions $\Phi(r)$ for which
(\ref{mass}) can be evaluated exactly is infinite. It should be
noted, however, that the generation of an exact solution does not
necessarily mean that the equation $p(r=R)=0$ can be solved
exactly. The algorithm presented here is now demonstrated by the
construction of an infinite number of previously unknown but
physically interesting exact solutions of Einstein's equations.

\bigskip
Let
\begin{equation}
\Phi(r)=\frac{1}{2}\,N\ln( 1+{\frac
{{r}^{2}}{{\alpha}}})\label{tolman}
\end{equation}
where $N$ is an integer $\geq1$ and $\alpha$ is a constant $>0$.
The function (\ref{tolman}) is monotone increasing with a regular
minimum at $r=0$. With the source function (\ref{tolman}),
(\ref{mass}) can be evaluated exactly for any $N$. Whereas
(\ref{tolman}) generates a ``class" of solutions, the metric (in
particular $m(r)$) looks quite distinct, and the physical
properties are quite distinct, for each value of $N$. Previously,
only for $N=1,...,5$ were solutions known, having been arrived at
by various methods, and one solution which is the first term in
the Taylor expansion of (\ref{tolman}) \cite{tolmaniv}. (These
solutions, with $N=1,...,5$, in fact constitute half of all the
previously known physically interesting solutions (of this type)
in curvature coordinates.)  For $N\geq5$ the solutions are
acceptable on physical grounds and even exhibit a monotonically
decreasing subluminal adiabatic sound speed \cite{new}.

\bigskip
It is, perhaps, worth noting here that the foregoing discussion in
curvature coordinates can be transformed directly into Bondi
radiation coordinates \cite{bondi}.

\bigskip
Now consider ``isotropic coordinates"
\begin{equation}
ds^2_{\mathcal{M}}=e^{2B(\textrm{r})}(d\textrm{r}^2+\textrm{r}^2d\Omega^2)-e^{2(\Psi(\textrm{r})-B(\textrm{r}))}dt^2.
\label{standardformiso}
\end{equation}
Unlike curvature coordinates, the isotropic form
(\ref{standardformiso}) does not offer an immediate invariant
physical interpretation of the functions $\Psi(\textrm{r})$ or
$B(\textrm{r})$ \cite{isometric}. However, as we now show, the
coordinates offer a simplified algorithm for constructing perfect
fluid solutions. Writing out (\ref{condition}) we now obtain an
expression involving $\Psi(\textrm{r})$ and $B(\textrm{r})$ with
derivatives to order two in $\Psi(\textrm{r})$ and to order one in
$B(\textrm{r})$. Viewing (\ref{condition}) as a differential
equation in $\Psi(\textrm{r})$, given $B(\textrm{r})$, we again
obtain a Riccati equation in the first derivative of
$\Psi(\textrm{r})$. However, viewing (\ref{condition}) as a
differential equation in $B(\textrm{r})$, given
$\Psi(\textrm{r})$, we obtain an equation solvable simply by
quadrature. As a consequence, we have the following simplified
algorithm for constructing all possible spherically symmetric
perfect fluid solutions of Einstein's equations in isotropic
coordinates:

\bigskip
Given $\Psi(\textrm{r})$ (sufficiently smooth and subject to
boundary conditions explained below)
\begin{equation}
B(\textrm{r}) =\Psi(\textrm{r})+\int c ( \textrm{r})
d\textrm{r}+\mathcal{C} \label{bsource}
\end{equation}
where
\begin{equation}
c(\textrm{r}) \equiv \frac{\epsilon}{\sqrt{2}}
\sqrt{(\Psi^{'}(\textrm{r}))^2-\Psi^{''}(\textrm{r})+\Psi^{'}(\textrm{r})/\textrm{r}}
 \label{isosource}
\end{equation}
with $\epsilon=\pm 1$, $^{'}\equiv\frac{d}{d\textrm{r}}$ and
$\mathcal{C}$ a constant. Recently, Rahman and Visser
\cite{visser} have also presented an algorithm for constructing
spherically symmetric perfect fluid solutions in isotropic
coordinates. The source function $\Psi(\textrm{r})$ used here is
related to the source function $z(\textrm{r})$ used by Rahman and
Visser as follows:
\begin{equation}
\Psi(\textrm{r})=2\int\frac{\textrm{r}
z(\textrm{r})}{1-z(\textrm{r})\textrm{r}^2}d\textrm{r}. \label{rv}
\end{equation}
The two algorithms differ fundamentally in the sense that only one
integration is used in the present procedure as opposed to two
distinct integrations used in the Rahman-Visser procedure. The
Rahman-Visser procedure was motivated by the requirement that
metric be manifestly real ab initio. The reality of the integral
(\ref{bsource}) is discussed below.

\bigskip
Interior boundary conditions on $\Psi(\textrm{r})$ are set exactly
as in the case of curvature coordinates. We now have the energy
density and pressure in the form
\begin{equation}
\rho=\frac{G1}{8 \pi}=\frac{-1}{8 \pi e^{2B(\textrm{r})}}(2
B^{''}(\textrm{r})+\frac{4
B^{'}(\textrm{r})}{\textrm{r}}+(B{'}(\textrm{r}))^2)
\geq0\label{energyiso}
\end{equation}
and
\begin{equation}
 p=\frac{G2}{8 \pi}=\frac{-1}{8 \pi e^{2B(\textrm{r})}}(-B{'}(\textrm{r})\Psi{'}(\textrm{r})+(B{'}(\textrm{r}))^2-2\frac{\Psi^{'}(\textrm{r})}{\textrm{r}})\geq0.\label{pressureiso}
\end{equation}
$\Psi(0)$ is a finite constant (set by the scale of $t$) and it
follows from (\ref{pressureiso}) that  $\Psi^{'}(0)=0$ and from
(\ref{bsource}) that $B^{'}(0)=0$. With $p(\textrm{r})\geq 0 $ it
follows that the source function $\Psi(\textrm{r})$ must be a
monotone increasing function with a regular minimum at
$\textrm{r}=0$ and $\Psi^{''}(0)=4 \pi e^{2B(0)} p(0)$. Exterior
boundary conditions on $\Psi(\textrm{r})$ are set as in curvature
coordinates. Regularity of $\rho(0)$ requires $B^{'}(0)=0$ and
with $\rho(\textrm{r})\geq 0 $ it follows that $B(\textrm{r})$
must be a monotone decreasing function with a regular maximum at
$\textrm{r}=0$ and $B^{''}(0)=-4 \pi e^{2B(0)} \rho(0)$. The
limits $-\frac{2}{\textrm{r}}<B^{'}(\textrm{r})<0$ guarantee the
positivity of the effective gravitational mass. To examine the
reality of the metric consider the function
$F(r)\equiv(\Psi^{'}(\textrm{r}))^2-\Psi^{''}(\textrm{r})+\Psi^{'}(\textrm{r})/\textrm{r}$.
Now $F(0)=0,\;F'(0)=0$ and $F''(0)>0$ so $F(r)$ has a local
minimum at $r=0$. Now suppose that $F(r)=0$ for $r>0$. Then
condition (\ref{condition}) requires $B^{'}=\Psi^{'}$ so we have
already passed through a region with $\rho<0$ before the reality
of the metric breaks down (in agreement with known theorems
\cite{br}).

\bigskip
In parallel to the algorithm in curvature coordinates, each source
function $\Psi(\textrm{r})$ which is a smooth monotone increasing
function with a regular minima at $\textrm{r}=0$ necessarily
gives, via (\ref{bsource}), a static spherically symmetric perfect
fluid solution of Einstein's equations which is regular at
$\textrm{r}=0$. Exact solutions are again those for which
(\ref{bsource}) can be evaluated without recourse to numerical
methods. Physical considerations must guide the choice of
$\Psi(\textrm{r})$. In isotropic coordinates, ratios of invariants
and differential invariants can be obtained directly from the
source function $\Psi(\textrm{r})$ via differentiation. You do not
need $B(\textrm{r})$ and in particular you do not need to
integrate. For example, the functions
$p(\textrm{r})/\rho(\textrm{r})$ and
$p^{'}(\textrm{r})/\rho^{'}(\textrm{r})$ follow directly without
integration. Of course, neither $p(\textrm{r})$ nor
$\rho(\textrm{r})$ follow without integration. In curvature
coordinates you cannot get these ratios without integration,
starting from the source function $\Phi(r)$.

\bigskip
To demonstrate the algorithm in isotropic coordinates let
\begin{equation}
\Psi(\textrm{\textrm{r}})=\alpha \ln
\frac{f(\textrm{r})}{g(\textrm{r})} \label{isogen}
\end{equation}
where $\alpha$ is a constant $>0$. Of course it is not difficult
to find functions $f(\textrm{r})$ and $g(\textrm{r})$ so that
(\ref{isogen}) is monotone increasing with a regular minimum at
$\textrm{r}=0$.  Nor indeed is it difficult to find such functions
for which $B(\textrm{r})$ can be evaluated exactly. For example,
let $g(\textrm{r})=(\delta+\epsilon \textrm{r}^2)^{\zeta}$ and
$f(\textrm{r})=\delta^{\zeta}+\gamma \textrm{r}^2$ with $\delta,
\epsilon, \gamma$ and $\zeta$ constants such that $\delta
> 0$ and $\delta^{1-\zeta} \gamma > \zeta \epsilon$. This class of
solutions includes a number of known solutions including the
Schwarzschild interior solution and the Rahman-Visser general
quadratic ansatz. Any solution in isotropic coordinates can be
immediately recovered and generalized following the algorithm
presented \cite{isosolutions}.

\bigskip
An algorithm based on the choice of a single monotone function
(subject to boundary conditions) has been presented which
generates all regular static spherically symmetric perfect fluid
solutions of Einstein's equations. In all cases the choice of
generating function must be guided by physical considerations.
These additional conditions limit the generating functions allowed
by way of non-trivial integral-differential inequalities. The
details of how to choose physically relevant generating functions
(beyond trial and error) are, at present, not known. Moreover, the
resultant equation of state is a byproduct of the algorithm and
can not be set a priori. Despite these reservations, the algorithm
has been demonstrated by the construction of an infinite number of
previously unknown physically interesting exact solutions
\cite{number}. It is a curious fact of history that over half a
century ago Max Wyman \cite{wyman} pointed out that the algorithm
presented here was possible and yet, despite the voluminous
literature on the subject \cite{dellake}, the algorithm appears
not have been followed up.

\begin{acknowledgments}
This work was supported by a grant from the Natural Sciences and
Engineering Research Council of Canada. Portions of this work were
made possible by use of \textit{GRTensorII} \cite{grt}. It is a
pleasure to thank Gyula Fodor, Jim Lattimer, Nicholas Neary, Don
Page and Matt Visser for comments and Jorge Pullin for pointing
out the paper by Berger, Hojman and Santamarina.
\end{acknowledgments}


\begin{thebibliography}{}\label{sec:TeXbooks}
\bibitem[*]{email}{Electronic Address: lake@astro.queensu.ca}
\bibitem{dellake}See M. S. R. Delgaty and K. Lake,
Computer Physics Communications  \textbf{115}, 395 (1998)
(gr-qc/9809013).
\bibitem{physical}The conditions used in \cite{dellake} were: (i) isotropy of the
pressure (otherwise any metric is a ``solution"), (ii) regularity
at the origin, (iii) positivity of the pressure and energy density
at the origin, (iv) vanishing of the pressure at a finite
boundary, (v) monotone decrease of the energy density to the
boundary and (vi) subluminal adiabatic sound speed. In addition to
these, a monotone decrease in the subluminal adiabatic sound speed
is desirable.
\bibitem{metric}We use geometrical units throughout. The
``curvature coordinates" used in (\ref{standardform}) have the
advantage that the metric functions have a clear invariant
physical interpretation (but see also \cite{bondi} below). The
function $m(r)$ is the effective gravitational mass. See W. C.
Hernandez and C. W. Misner, Astrophys. J. \textbf{143}, 452
(1965), E. Poisson and W. Israel, Phys. Rev D \textbf{41}, 1796
(1990), T. Zannias, Phys. Rev. D \textbf{41}, 3252 (1990) and S.
Hayward Phys.Rev. D {\bf 53}, 1938 (1996) (gr-qc/9408002). Whereas
$\Phi(r)$ is (in the weak field limit) the ``Newtonian" potential,
$re^{-\Phi(r)}$ is the effective potential for null geodescics
(see, for example, M. Ishak, L. Chamandy, N. Neary and K. Lake,
Phys. Rev. \textbf{D64} 024005 (2001) (gr-qc/0007073)).
\bibitem{lake}gr-qc/0209063
\bibitem{TOV} One can take the view that the Tolman-Oppenheimer-Volkoff
equation is a conseuence of the invariant statement
(\ref{condition}).
\bibitem{condition} Explicitly, condition (\ref{condition}) in the static case in curvature coordinates
reduces to the Walker pressure isotropy condition
$G^{r}_{r}=G^{\theta}_{\theta}$ (see A. G. Walker, Quarterly
Journal of Mathematics, \textbf{6}, 81 (1935)) which is
\begin{widetext}
\begin{eqnarray}
 \left( {\frac {d^{2}}{d{r}^{2}}}\Phi \left( r \right) + \left( {
\frac {d}{dr}}\Phi \left( r \right)  \right) ^{2} \right) {r}^{2}
 \left( r-2\,m \left( r \right)  \right) -r \left( {\frac {d}{dr}}\Phi
 \left( r \right)  \right)  \left(  \left( {\frac {d}{dr}}m \left( r
 \right)  \right) r+r-3\,m \left( r \right)  \right) +3\,m \left( r
 \right) - \left( {\frac {d}{dr}}m \left( r \right)  \right)
 r=0. \nonumber
\end{eqnarray}
\end{widetext}
\bibitem{berger} The problem has also been reduced to a
linear equation of first order by A. S. Berger, R. Hojman and J.
Santamarina, J. M. P. \textbf{28}, 2949 (1987). Recently G. Fodor
(gr-qc/0011040) has reduced the problem to an algebraic one with
integration required only for one metric function but not the
physical variables $\rho$ and $p$.
\bibitem{invariants}D. Pollney, N. Pelavas, P. Musgrave and K. Lake,
Computer Physics Communications  \textbf{115}, 381 (1998)
(gr-qc/9809012).
\bibitem{schw} It follows from (\ref{condition}) and (\ref{standardform}) that the necessary and
sufficient condition for conformal flatness for $r>0$ is given by
$m(r)=cr^3$, which gives, uniquely, the Schwarzschild interior
solution. See also H. A. Buchdahl, A. J. P. \textbf{39}, 158
(1971).
\bibitem{br} See T. W. Baumgarte and A. D. Rendall, Class. Quant.
Grav. \textbf{10}, 327 (1993) and also M. Mars, M. Merc\`{e}
Mart\'{i}n-Prats and J. M. M. Senovilla, Phys. Lett \textbf{A
218}, 147 (1996) (gr-qc/0202003).
\bibitem{junctions} See, for example, P. Musgrave and K. Lake,
Class. Quantum Grav. \textbf{13}, 1885 (1996) (gr-qc/9510052). At
an interior boundary surface $p$, but not $\rho$, must be
continuous. Discontinuities in $\rho$ are associated with phase
transitions which we do not considered here. For a discussion of
interior phase transitions see, for example, L. Lindblom, Phys.
Rev. \textbf{D 58}, 024008 (1998).
\bibitem{tolmaniv} In terms of the
classification given in \cite{dellake} the solutions are
\textbf{Tolman IV} for $N=1$, \textbf{Heint IIa} for $N=3$,
\textbf{Durg IV} for $N=4$ and \textbf{Durg V=D-F} for $N=5$. If
$\Phi(r)$ is taken to be the first term in the Taylor exapnsion of
(\ref{tolman}), the solution is known as \textbf{Kuch2 III}. The
case $N=2$  gives $m(r)= \mathcal{C}{r}^{3}/( 3 r^{2}+\alpha) ^{2/
3} $ which is usually dismissed under the erroneous assumption
$\mathcal{C}=0$.
\bibitem{new} N. Neary, J. Lattimer and K. Lake (in preparation).
\bibitem{bondi} These were first discussed (in the spherically symmetric case)
by  H. Bondi, Proc. R. Soc. London \textbf{A281}, 39 (1964) and
are a generalization of the well known Eddington-Finkelstein
coordinates for the Schwarzschild vacuum.  The algorithm presented
is equally at home in curvature and radiation coordinates. Writing
\begin{eqnarray}
dv=dt \pm \frac{e^{-\Phi(r)}}{\sqrt{1-2m(r)/r}}\; dr, \nonumber
\end{eqnarray}
$+$ for advanced (ingoing) $v$ and $-$ for retarded (outgoing) $v$
it follows that (\ref{standardform}) takes the form
\begin{eqnarray}
ds^2 = \pm 2 \frac{e^{\Phi(r)}}{\sqrt{1-2m(r)/r}}\;dv dr+r^2 d
\Omega^2 -e^{2\Phi(r)} d v^2. \nonumber
\end{eqnarray}
The form of condition (\ref{condition}) (given above in \cite
{condition}) remains unchanged as do the functional forms and
physical meanings of $\Phi,\; m,\; \rho$ and $p$.
\bibitem{isometric}We proceed here in isotropic
coordinates ab initio without coordinate transformations. Now
$re^{2B(\textrm{r})-\Psi(\textrm{r})}$ is the effective potential
for null geodescics and the effective gravitational mass is given
by $m(\textrm{r})=-(B^{'}(\textrm{r})(B^{'}(\textrm{r})
\textrm{r}+2) e^{B(\textrm{r})}\textrm{r}^2 /2$ where $^{'} \equiv
d/d\textrm{r}$.
\bibitem{visser}S. Rahman and  M. Visser, Class. Quant. Grav. \textbf{19},  935
(2002) (gr-qc/0103065).
\bibitem{isosolutions} In the terminology of \cite{dellake}, as regards physically interesting solutions, the \textbf{P-S2}
solution follows from the stated form of $\Psi(\textrm{r})$.
Similarly, the choices $g(\textrm{r})=cosh(\beta+\gamma r^2)$ and
$f(\textrm{r})=sinh(\beta+\gamma r^2)$ with $\beta$ and $\gamma$
positive constants immediately gives \textbf{Gold III}.
\bibitem{number} It is also of interest to note that seven of the
eleven previously known solutions of this type are special cases
resulting from the two generating functions considered here.
\bibitem{wyman}M. Wyman, Phys. Rev \textbf{75}, 1930 (1949).
\bibitem{grt}This is a package which runs within Maple. It is entirely
distinct from packages distributed with Maple and must be obtained
independently. The GRTensorII software and documentation is
distributed freely on the World-Wide-Web from the address \textit{
http://grtensor.org}
\end{thebibliography}
\end{document}